\documentclass[preprint]{aastex}
\begin{document}
\title{Soft X-ray spectroscopy of highly charged silicon ions in dense plasmas}
\author{G.Y. Liang\altaffilmark{1}, G. Zhao\altaffilmark{1}, and J.Y. Zhong\altaffilmark{1,2}}
\affil{National Astronomical Observatories, Chinese Academy of
Sciences} \email{gzhao@bao.ac.cn} \and
\author{Y.T. Li\altaffilmark{2}, Y.Q. Liu\altaffilmark{2}, Q.L. Dong\altaffilmark{2}, X.H. Yuan\altaffilmark{2}, Z. Jin\altaffilmark{2}, and J. Zhang\altaffilmark{2}}
\affil{Institute of Physics, Chinese Academy of Sciences}
\email{jzhang@aphy.iphy.ac.cn}

\altaffiltext{1}{A20 Datun Road, Chaoyang District, Beijing
100012, China.} \altaffiltext{2}{P.O.Box 603, Beijing 100080,
China.}

\begin{abstract} Rich soft X-ray
emission lines of highly charged silicon ions (Si~VI--Si~XII) were
observed by irradiating an ultra-intense laser pulse with width of
200~fs and energy of $\sim$90~mJ on the solid silicon target. The
high resolution spectra of highly charged silicon ions with
full-width at half maximum (FWHM) of $\sim$0.3--0.4\AA\, is
analyzed in wavelength range of 40--90~\AA\,. The wavelengths of
53 prominent lines are determined with statistical uncertainties
being up to 0.005~\AA\,. Collisional-radiative models were
constructed for Si~VI -- Si~XII ions, which satisfactorily
reproduces the experimental spectra, and helps the line
identification. Calculations at different electron densities
reveal that the spectra of dense plasmas are more complicate than
the spectra of thin plasmas. A comparison with the Kelly database
reveals a good agreement for most peak intensities, and
differences for a few emission lines.
\end{abstract}


\section{Introduction}
Spectroscopic observations with the space-based observatories in
soft X-ray and extreme-ultraviolet (EUV) regions provide valuable
diagnostic means for understandings of the astrophysical hot
plasma. The soft X-ray and EUV regions contain rich emission lines
that can be used for determinations of plasma properties and
element abundances over a wide temperature
range~\citep[etc.]{MRD01,Ness01,NSB02,NSBM02,NBD03}. The emissions
are mainly from L-shell transitions of abundant astrophysical
elements and M-shell transitions of iron ions. The most abundant
element--Fe, shows a forest-like emission lines at the soft X-ray
and EUV regions in the solar and stellar observations. So the
spectral analyses of L-shell iron ions received strong attentions
for astronomers and physicians. However, the emissions lines
mainly span 6--18~\AA\, ($\Delta n \geq 1$, here $n$ is main
quantum number) and 100--350~\AA\, ($\Delta n~=~0$) regions. So
previous space observatories, e.g. {\it Skylab} (covering
6--49~\AA\,), Solar EUV Rocket Telescope and Spectrograph (SERTS,
170--450~\AA\,), Extreme Ultraviolet Explore ({\it EUVE}, covering
70--760~\AA\, with a resolution of
$\Delta\lambda$=0.4--2.5~\AA\,)\citep{BDE00}, and Solar
Ultraviolet Measurements of Emitted Radiation (SUMER, covering
500--1610~\AA\, with resolution power of
$\lambda/\Delta\lambda$~=~19000--40000) instrument on SOHO mission
etc., probe into the universe in the two wavelength ranges.
However, the spectra between 40--90~\AA\, has received a scant
attention even in solar measurements. Though the new generation of
space mission of {\it Chandra X-ray Observatory} covers wavelength
range of 1.2--176\AA\, with higher resolution of
$\sim$0.06\AA\,\citep{BGK00,MRD01,NSBM02,AMP03}, most researches
focus on observations below $\sim$35\AA\, provided by High Energy
Transmission Grating Spectrometer (HETGS) (with
$\Delta\lambda$=0.012 and 0.023\AA\,) on {\it Chandra} and
Reflected Grating Spectrometer (RGS) (with
$\Delta\lambda$=0.05\AA\,) on {\it XMM-Newton}
\citep{CHD00,ABG01,BEG01}. This is due to the spectral region is
spanned by $\Delta n\geq 1$ transition lines of highly charge iron
ions.

The region of 40--90\AA\, contains more lines than currently
identified, as graphically illustrated by {\it Chandra} spectra of
Capella \citep{BGK00} and Procyon (Raassen et al. 2002) observed
with Low Energy Transmission Grating Spectrometer (LETGS). Those
unresolvable weak lines form the pseudo-continuum emissions, and
raise the ``background" level for strong features, and increase
the uncertainty to their interpretation. Contribution to these
unidentified lines may come from L-shell ions of magnesium,
silicon, sulfur, argon, calcium, iron and nickel. Nevertheless,
the uncertainties of wavelength and cross section of excitation,
ionization and recombination are more significant for L-shell ions
than K-shell ions of above listed elements. The associated
uncertainties greatly impede our efforts to fully understand
astrophysical X-ray sources, such as the shape of emission measure
distribution, element abundances, and spatial structures. So
laboratory studies, using, for example, electron-beam ion traps
(EBIT), storage rings, tokamaks, and intense laser pulses, are
very necessary to provide accurate atomic data that can be
incorporated into astrophysical spectral synthesis codes.

The interaction of intense, sub-picosecond laser pulses with solid
targets allows the generation of hot plasmas with a temperature
between 50~eV and 1~keV, which is the typical temperature range of
stellar coronae, and a density of
$10\time10^{19}$--$10\time10^{21}$~cm$^{-3}$. The laser plasma is
one of ground X-ray and EUV sources. This type plasma plays an
important role in various fields of research such as astrophysics,
inertial confinement fusion, etc (Lindl et al. 2004). Its X-ray
and EUV radiation can be used to determine plasma conditions and
micro-processes. For example the density can be obtained from line
broadening effects, and the temperature can be estimated from
satellites of H-like Ly$\alpha$ and He-like resonance
(1s2p~$^1$P$_1$--1s$^2$~$^1$S$_0$) lines. The well controlled
laser plasma can also be used to benchmark the kinetic codes
\citep{GFD00}, and simulate some astrophysical phenomena.

Recently, many efforts have been made in laboratories to provide
accurate atomic data and validate the simulation codes.
\cite{GFD00} and other groups \citep{SFD97,BRF02} investigated the
X-ray emissions of K-shell ions of argon, using the intense pulsed
laser, electron beam ion traps (EBIT) and tokamaks. Independent
methods such as Thomson Scattering and infrared interferometer
measurement for temperature and density, have been used to
benchmark the $T_{\rm e}$ and $n_{\rm e}$ constrained by
spectroscopy, and a good agreement is obtained. Lepson et al.
(2003, 2005) further investigated the soft X-ray emissions of
L-shell ions of argon (20--50~\AA\,) and sulfur (20--75\AA\,)
using the Lawrence Livermore EBIT-I and EBIT-II, and graphically
illustrated larger uncertainties in available atomic data.
\cite{YMW00} measured emissions of sulfur in wavelength range of
190--400~\AA\, by beam-foil spectroscopy in the Heavy Ion Research
Facility at Lanzhou.

In this paper, the experimental setup and the calibration for the
soft X-ray and EUV spectrometer are described in Sect.~2. Results
and discussions are presented in Sect.~3. In this section, the
soft X-ray spectra of highly charged silicon ions, modelling to
the spectra, and the comparison between the present calculation
and the Kelly database are described in detail. Finally, the
conclusions is outlined in Sect.~4.

\section{Experimental setup and spectrometer calibration}
The experiment was carried out using the Xtreme Light II (XL-II)
laser system at the Institute of Physics (IoP), Chinese Academy of
Sciences (Peng et al. 2004, Li et al. 2006). The laser system has
an output energy up to 650mJ in 30fs pulses at 796~nm. The laser
beam was focused by an $f/5$ lens to a dimension of
$\sim$26$\times$24~$\mu$m on the surface of a planner target,
which is consist of three materials including copper, plastic and
silicon as shown by the left-top panel of Fig.~1. The copper is
used for collimation, while the plastic is used for wavelength
calibration in the selected wavelength range. A grazing-incidence
flat field grating spectrometer with varied spacing grating
(average groove density is 1200~$l$/mm), was adopted to disperse
the radiation. A cooled CCD camera (running at -24~$^{\circ}$C) is
used for readout of the spectrograph. The CCD chip has
2048$\times$512~pixels with each of
$\sim$13.5$\times$13.5~$\mu$m$^2$ and thus, the CCD total surface
area is about 28$\times$7~mm$^2$. The detail description refers to
the work of Liu et al. (2004). The spectrometer can cover soft
X-ray and Extreme Ultraviolet (EUV) region (40--400~\AA\,) by
sliding the CCD camera along the dispersion plane. In the present
single setting, the wavelength coverage is 40--180~\AA\,, which is
the shortest wavelength range covered by the designed
spectrometer. The spectral resolution is about
$\delta\lambda\sim$0.3 over 40--90~\AA\,.

Figure 1 shows the experimental layout. The laser beam with energy
of 90~mJ and pulse duration of 200~fs was focused onto the target
in direction of 45$^{\circ}$ to the target normal, with intensity
of $\sim10^{15}$~W/cm$^2$. Firstly, the focused laser irradiates
the plastic. As shown by the CCD spectral image in Fig.1, there is
a slight rotation between pixel rows and the dispersion direction.
A program is compiled to correct this rotational misalignment
before summing over the pixel channels. Additionally, a clear
discrete band is observed in direction being perpendicular to the
grating dispersion. This is due to the toroidal mirror placed
before the entrance slit. So the spectra in pixel region of
60--260 is selected in summing over the pixel channels.

In order to obtain the dispersion function of the present
spectrometer, 14 well-known lines from highly charged carbon,
nitrogen and oxygen ions, and their higher order diffraction
covering the whole wavelength range of interest are used, as shown
by symbols in Fig.~2. Peak centroids of these emission lines are
determined through Gaussian fit, which can describe the observed
line shape satisfactorily as shown by the inset in Fig.~2. A
full-width at half maximum (FWHM) of $\sim$3.5--4.5 pixels is
found, corresponding to $\delta\lambda~\sim$0.3--0.4~\AA\,. The
measured line width ($\sim$3.5$\times$13.5=47~$\mu$m, with each
pixel size of 13.5~$\mu$m) reflects the width of entrance slit of
50~$\mu$m. The wavelength values of the calibration lines were
adopted from NIST
database\footnote{http://physics.nist.gov/PhysRefData/ASD/index.html}
and the website {\sf
http://www.pa.uky.edu/$^{\sim}$peter/atomic/}. A parabolic
polynomial is used to represent the pixel ($x$)-wavelength
($\lambda$) relation [see Fig.~2]. The calibration uncertainty is
estimated to be 0.11~\AA\, [rms value] over the range of
40--180~\AA\,. Multi-measurements and calibrations reveal that the
calibration parameters hold constant.

\section{Results and discussions}
\subsection{Soft X-ray spectrum of highly charge silicon}
When the focused laser irradiate on the solid target, rich
emission lines are observed, as shown in below image of Fig.~1. In
this work, we pay special attention on the soft X-ray emission
spectra in the wavelength range of 40--90~\AA\,, as labelled out
by white region in Fig.~1. 10 separate measurements were performed
with each measurement of 4 continuous laser pulses irradiating on
the target and 30~s accumulation. Using the calibration
parameters, we scale all the measurements. Fig.~3 illustrates the
averaged spectrum and its best-fitting in wavelength range of
40--90~\AA\,. In this region, emission lines are dominated by
$\Delta n \geq 1$ transition lines of Si~VI--Si~XII ions.

We separately analyze the 10 spectra, and accurately determine the
wavelengths for 53 prominent lines. A statistical accuracy of
0.005~\AA\, is obtained for those strong and blend-free lines, as
listed in the second column of Table 1.

Based upon the theoretical wavelengths and relative line
intensities [explained in the next subsection], we identified
these emission lines. The identified wavelength values are
partially from our recent calculations~\citep{LZZ07}, NIST
database\footnote{http://physics.nist.gov/PhysRefData/ASD/index.html}
and Kelly
database\footnote{http://cfa-www.harvard.edu/amp/ampdata/kelly/kelly.html}~\citep{Kel87}.
In the third subsection, the detailed comparison of the present
calculation and the Kelly database is given. From Table 1, the
differences between the measurements and available database are
within 0.1~\AA\, for most emission lines, whereas differences up
to 0.3~\AA\, also appears for some emission lines. This reveals
that detailed structure calculations for L-shell silicon ions by
inclusion of large configuration interactions and relativistic
effects are necessary.

\subsection{Modelling}
In soft X-ray region, forest-like lines from L-shell ions
generally group together or blend each other when the broadening
from spectrometers has been considered, as in this case. Besides,
large discrepancies of line wavelengths between measurements and
calculations have been demonstrated for highly charged sulfur
($\sim$0.3~\AA\,) and argon ($\sim$0.1~\AA\,) ions. So the
predictions of line intensities are necessary for the line
identification.

Collisional-radiative (CR) models for highly charged Si~VI--Si~XII
ions are constructed to simulate the experimental spectrum. The
atomic data used to calculate the line intensities, are generated
with the Flexible Atomic Code (FAC) provided by \cite{Gu03}. A
fully relativistic approach based on Dirac equation is used
throughout the entire package. Energy levels, transition rates of
E1, E2, M1, and M2 types, and electron impact excitation strengths
have been revaluated for Si~IX--Si~XI in our recent work
\citep{LZZ07}. 782, 878, 312, 560, 320, 350 and 40 energy levels,
have been included in predictions of line intensities for
Si~VI--Si~XII, respectively. These levels belongs to not only
singly excited configurations, but also some doubly excited
configurations for accounting for configuration interaction effect
as fully as possible. All possible $\Delta n=0$ (2--2) and $\Delta
n>0$ (2--3, 2--4, and 2--5) decay rates, excitations and
de-excitations among levels mentioned above have been considered
in the present simulation work.

The theoretical spectra have been computed in the steady state
equilibrium. The set of CR equations are solved by normalizing the
total level populations of each charge state to 1. The theoretical
spectrum for each charge stage is calculated at a density of
1.0$\times$10$^{20}$cm$^{-3}$ and temperatures of maximum fraction
in ionization equilibrium of Mazzotta et al. (1998). The
calculated line intensities are fold by Gaussian profile with FWHM
of 0.3~\AA\, being comparable to the experimental resolution. The
theoretical spectra are normalized by the measured values at
80.353~\AA\,, 71.425~\AA\,, 62.052~\AA\,, 55.569~\AA\,,
50.278~\AA\,, 47.670~\AA\, and 44.135~\AA\, for Si~VI--Si~XII,
respectively. As shown in Fig.~4, the simulation satisfactorily
reproduces the experimental spectra of highly charged silicon
ions. Different color curves indicate the spectra of different
charge states. By the comparison, most prominent emission lines
are identified as given in Table 1.

Since the broadening from the entrance slit, the measured line
width is about 0.3--0.4~\AA\,. Many emission lines of L-shell ions
grouped together, as illustrated by the spectra of Si~XI (dashed
and/or dotted curves) in Fig.~5-(a), so that the line
identification is very difficult. In thus case, we usually assign
the largest contribution to the emission peak. For example, we
assign the peaks at 46.403 and 47.670~\AA\, to the
$2s3d~^3D_2$--$2s2p~^3P_2$ (46.399~\AA\,) and
$2p3d~^3D_3$--$2p^2~^3P_2$ (47.653~\AA\,) transitions of Si~XI,
respectively.

It should be noted that at the density of the laser plasma, more
than half an ion's total level population can be in excited
levels, and even in doubly excited levels, other than the ground
level. So many lines excited from lowest lying excited levels can
be observed, e.g. lines at 46.403~\AA\, of Si~XI, 49.828~\AA\, of
Si~X, 55.950~\AA\, of Si~IX etc. Due to the high-density effect,
the spectra of laser plasma is more complicate than the spectra in
low density plasmas such as the astrophysical, electron beam ion
trap, and tokamak plasmas. In Fig.~5-(b), we show the spectrum of
Si~XI at high (1.0$\times$10$^{20}$cm$^{-3}$, blue-solid curve)
and low (1.0$\times$10$^{10}$cm$^{-3}$, red-solid curve) electron
densities. The emission line intensities around 47.67~\AA\, are
greatly suppressed, whereas the line intensities around
46.403~\AA\, are enhanced at the low-density plasma. This is due
to that the emission lines around 47.67~\AA\, are from higher
excited level of $2p^2$ configuration. In case of the low-density
plasma, the level population is mainly in the ground level.

\subsection{Comparison between present calculation and Kelly database}
Based upon the normalization for each charge states [see caption
of Fig.~4], we compare present spectra with the results from Kelly
database~\citep{Kel87}. For lines at 46.403 and 49.343~\AA\, of
Si~XI, present calculation shows a better agreement with the
measurement [see Fig.~4]. Moreover, Kelly predicted a stronger
Si~XI line at 50.517~\AA\, than the present calculation, which
blends with an emission line at 50.278~\AA\, of Si~X. By
multi-Gaussian component fitting, contribution from the two
features can be estimated in the measurement at the peak position
of 50.50~\AA\, as listed in Table 1. The blending results into two
peaks around 50.5~\AA\, in the sum spectrum [see Fig.~6] of Kelly.
Another strong Si~XI line at 52.296~\AA\, is predicted in Kelly
work, which helps our identification for the peak at 52.291~\AA\,
in the measurement. However, this emission line is missed in our
calculation.

For peak at 51.796~\AA\,, present calculation is higher than the
measurement by 30\%, whereas Kelly database misses lines at this
peak position. For the peak at 53.736~\AA\,, the two predictions
agree with each other, and well agree with the measurements [see
Fig.~4]. Kelly predicts a strong line at 54.599~\AA\, of Si~X,
which is two times of the present prediction. The present value is
comparable with the measured intensity at 54.705~\AA\,. So we
assign the emission line to Si~X (54.599~\AA\,).

For peak around 55.034, 55.379, 55.567, and 55.950~\AA\,, present
calculation predicts 4 Si~IX lines with intensities being
agreement with the measurement. Contributions from a group weak
Si~X lines (around 55.969~\AA\,) compensates the difference
between experiment and prediction of Si~IX at the peak of
55.950~\AA\,. Kelly database further predicts a Si~X line
(55.096~\AA\,) at the peak of 55.034~\AA\,, which results a large
deviation at the peak at 55.034~\AA\,. For the peak at 55.567 and
55.950~\AA\,, Kelly's values are lower than the measurement by
$\sim$40\%, and no other weak blending. For the peak at
59.088~\AA\, Kelly and this work predict a strong line at
59.004~\AA\,. Another important contribution from Si~IX line at
58.906~\AA\, is estimated in this work, whereas Kelly predicts
another line of Si~VIII (58.885~\AA\,) being higher than present
value by factors. The two work predict a strong Si~VII line close
to the peak position (59.088~\AA\,), which results into the large
difference between the sum spectra and the measurement at this
wavelength [see Fig.~6].

For the peaks at 61.256 and 62.052~\AA\,, Kelly and present works
predict two Si~VIII lines (see Table 1) with intensities being
agreement with the measurements [see Fig.~4]. However, Kelly
further shows a considerable contribution from Si~X and Si~VII at
the peak of 61.256~\AA\,, which results into the sum intensity is
higher than the experiment value by $\sim$60\%. For peaks at
62.993, 63.377, 63.762, and 64.482~\AA\,, Kelly's predictions show
a better agreement with the measurement than the present
calculation. Whereas, the peak intensities at 65.977 and
67.418~\AA\, are satisfactorily reproduced in the present work,
yet Kelly misses the contribution around 65.977~\AA\, and
overestimates the contribution around 67.418~\AA\, and
72.389~\AA\, by a factor up to $\sim$1.0.

For the peaks at 68.294, 69.734, 70.129, and 71.953~\AA\,, the two
calculations predict line intensities of Si~VII being consistent
with the measurements [see Fig.~4]. However, considerable
contribution from Si~VIII (69.790~\AA\,
$2s^22p^2(^3p)3s~^4P_{3/2}$--$2s^22p^3~^4S_{3/2}$) is predicted in
Kelly database, which results into the sum intensity is greatly
higher than the measured value. For the peaks at 73.136, 73.394
and 75.160~\AA\,, the two works satisfactorily reproduce the
measurement, whereas a considerable contribution from Si~VI is
estimated in Kelly database at 75.160~\AA\,. For the peaks 79.132,
79.448, 81.382, 81.698 and 83.809~\AA\,, present calculation
underestimates the contribution from Si~VII, however Kelly
database overestimates the contribution. Additionally, Kelly
database satisfactorily reproduce the peak at 84.935 and
85.330~\AA\, with wavelength higher than the measurement by
$\sim$0.28~\AA\,. But present prediction greatly underestimate the
line intensities. Around the peak at 88.0~\AA\,, one strong
emission line with wavelength of 87.641 and 88.008~\AA\, is
predicted by present model and the Kelly database, and the Kelly
database shows a better agreement with the measured value. The
partially blended emission line is from the second order
diffraction of the Si~XII line at 44.165~\AA\,.

For the emission lines of Si~VI ions, Kelly's result shows a good
agreement with the measurement at 77.348 and 77.626~\AA\,, and
better than the present prediction. For the peaks at 83.328 and
83.808~\AA\,, our results agree with the Kelly predictions,
whereas they are lower than the measured values by
$\sim$30\%--90\%.

\section{Conclusions}
Soft X-ray spectroscopy of highly charged silicon has been
measured with the flat-field grating spectrometer mounted on one
observation port of the Xtreme Light II at the Institute of
Physics, Chinese Academy of Sciences. The observed line-width is
$\sim$0.3--0.4~\AA\, over the wavelength range of 40--180~\AA\,,
which is due to the broadening from entrance slit with width of
50~$\mu$m. Using well-known lines of highly charged carbon,
nitrogen and oxygen, the dispersion function of the spectrometer
is calibrated with calibration uncertainty of 0.11~\AA\, over the
wavelength range of 40--180~\AA\,. In multi-measurements, 53
prominent lines are reproduced and their wavelengths are
determined with accuracy of 0.11--0.15~\AA\, (including
statistical uncertainty listed in Table~1, and the calibration
uncertainty of 0.1~\AA\,).

Collisional-radiative models has been constructed for each charge
states which satisfactorily reproduce the measurements. With the
aid of the simulation, we identified the 53 prominent lines. The
wavelength differences between present calculations and
measurements is within 0.1~\AA\, for most lines. In some case, the
differences can be up to 0.3~\AA\,, which reveals the detailed
atomic structure calculation with inclusion of large configuration
interaction and relativistic effects, is necessary.

Calculations at different electron densities, indicate that the
spectra of the dense plasma are more complicate than the spectra
from thin plasmas such as astrophysical, electron beam ion trap
and tokamak plasmas. This is due to that the high density effect
results into the higher level population in the excited levels
than in cases of the low-density plasmas.

A comparison with the Kelly database is also performed, which
reveals a good agreement for most peaks, and differences for some
peak intensities, such as at 71.953, 77.348, 81.698, 83.809, and
87.629~\AA\,. Similarly, the two predictions are benchmark by
comparing with the measurements. The presence of discrepancies
also leave a scope to investigate the density effect in the
high-dense plasma.

\acknowledgements This work was supported by the National Natural
Science Foundation of China under Grant No. 10603007, 10510490,
and 10521001, as well as National Basic Research Program of China
(973 Program) under grant No. 2007CB815103.

%figure-1
\clearpage
\begin{figure*}[h]
\centering
\includegraphics[angle=0,width=12cm,clip]{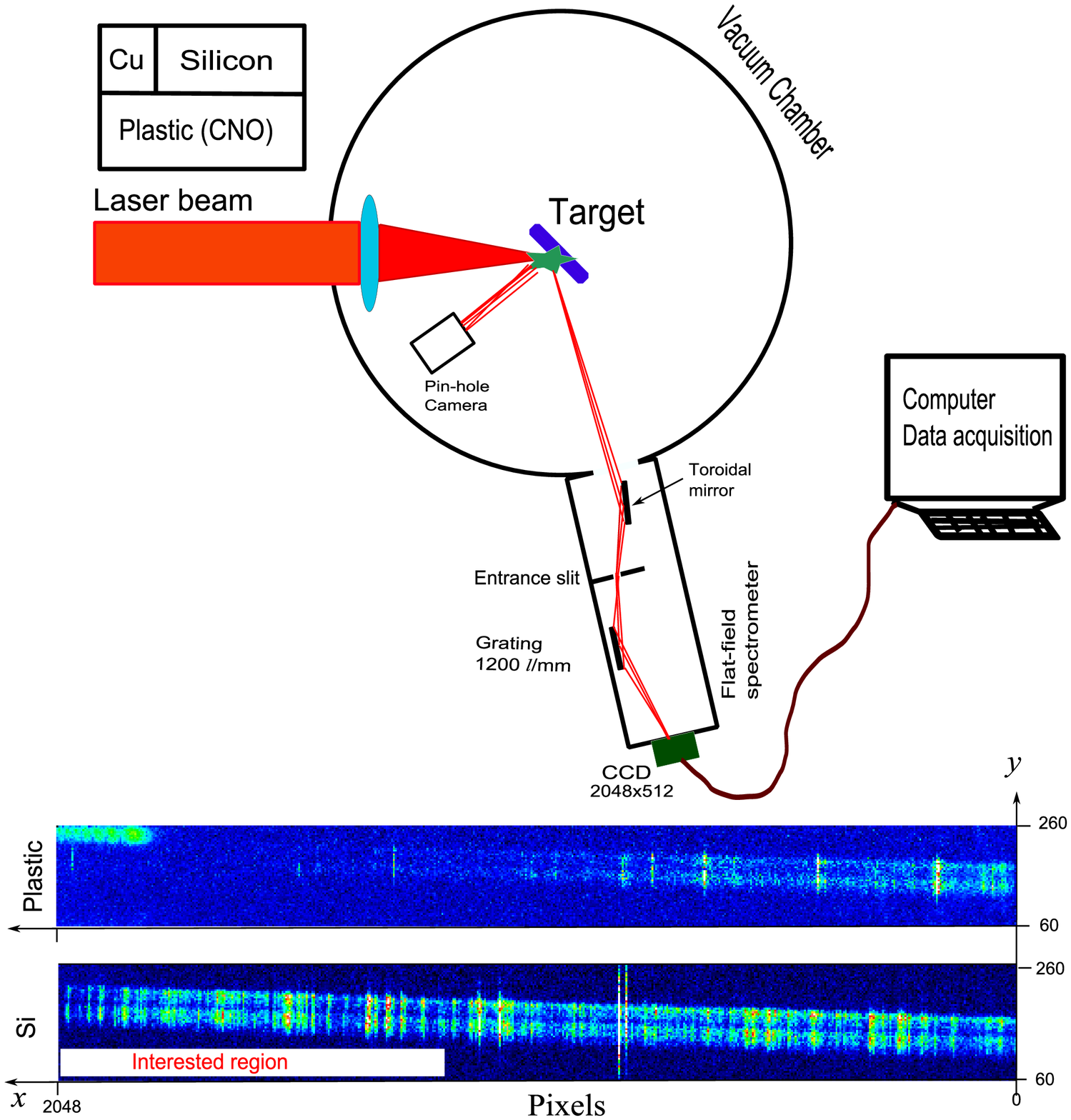}
\caption[short title]{(Online color)~~The schematic diagram of the
experimental setup. The left-top panel is the target
configuration. The bottom panel is partial readout image
(60--260~pixel region in direction of perpendicular to dispersion
direction) of the CCD camera. The interest region of this work is
denoted by the white stray.}
\end{figure*}
\clearpage

%figure-2
\clearpage
\begin{figure*}[h]
\centering
\includegraphics[angle=0,width=14cm,clip]{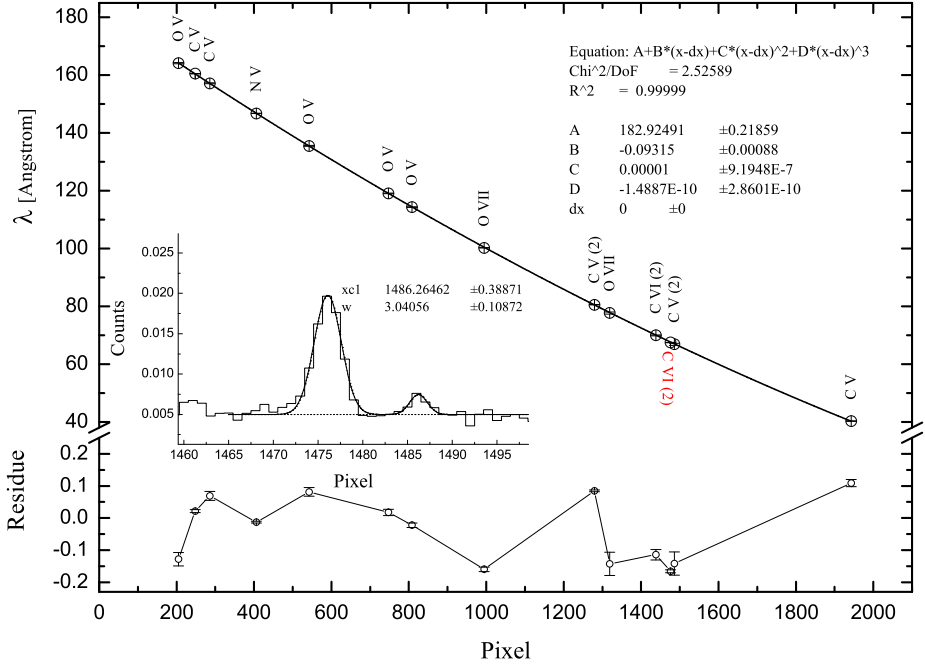}
\caption[short title]{(Online color)~~Dispersion function of the
flat-field soft X-ray/EUV spectrometer obtained from a cubic
polynomial fit and its residue (bottom) in wavelength range of
40--180~\AA\,. Symbols with error bars are calibration lines and
their residues. The inset spectrum shows a calibration line used
and its fit with Gaussian profile, which is marked by red
``C~VI~(2)" around dispersion curve.}
\end{figure*}
\clearpage

%figure-3
\clearpage
\begin{figure*}[h]\centering
\includegraphics[angle=0,width=15cm]{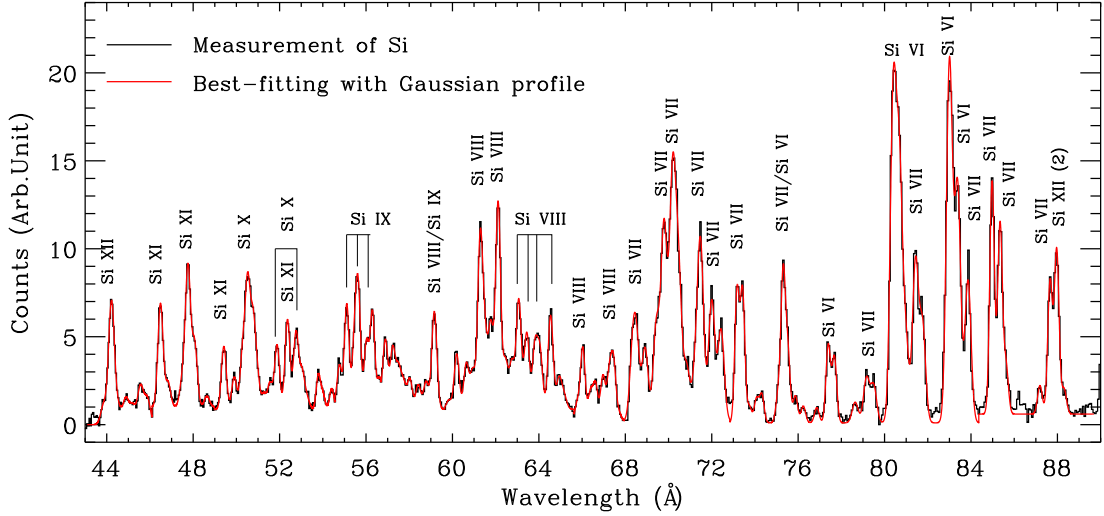}
\caption[short title]{(Online color)~~Averaged spectrum (black
histogram) of 10 measurements of highly charged silicon ions
produced by Xtrem Light II laser system (with a pulse duration of
$\sim$200~fs and an energy of $\sim$90~mJ) at the IoP, and its
best-fitting (red) in range of 40--90~\AA\,. For prominent lines,
identifications are labelled out.}
\end{figure*}
\clearpage

%figure-4
\clearpage
\begin{figure*}[h] \centering
\includegraphics[angle=0,height=10cm,clip]{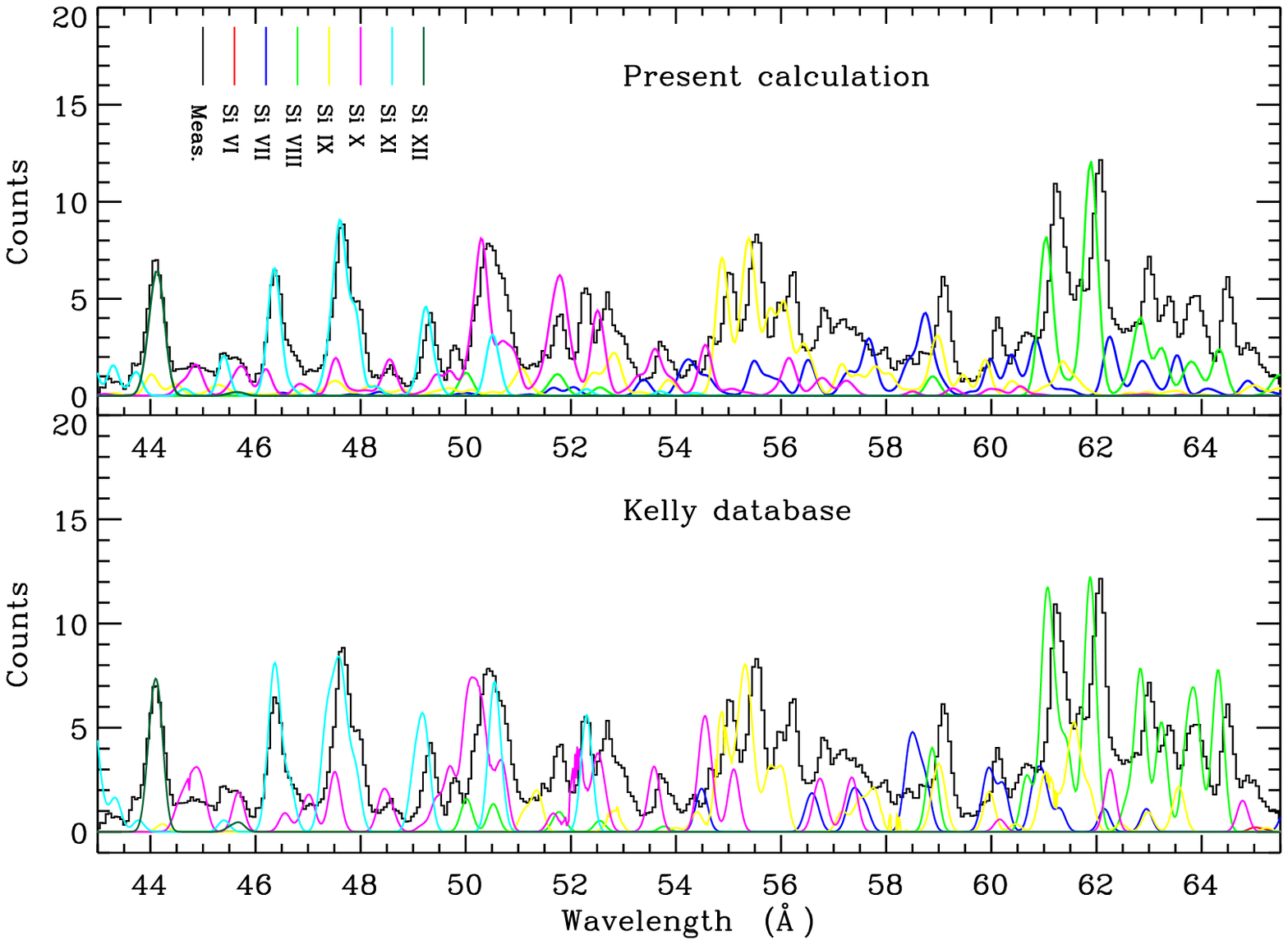}\\
\includegraphics[angle=0,height=10cm,clip]{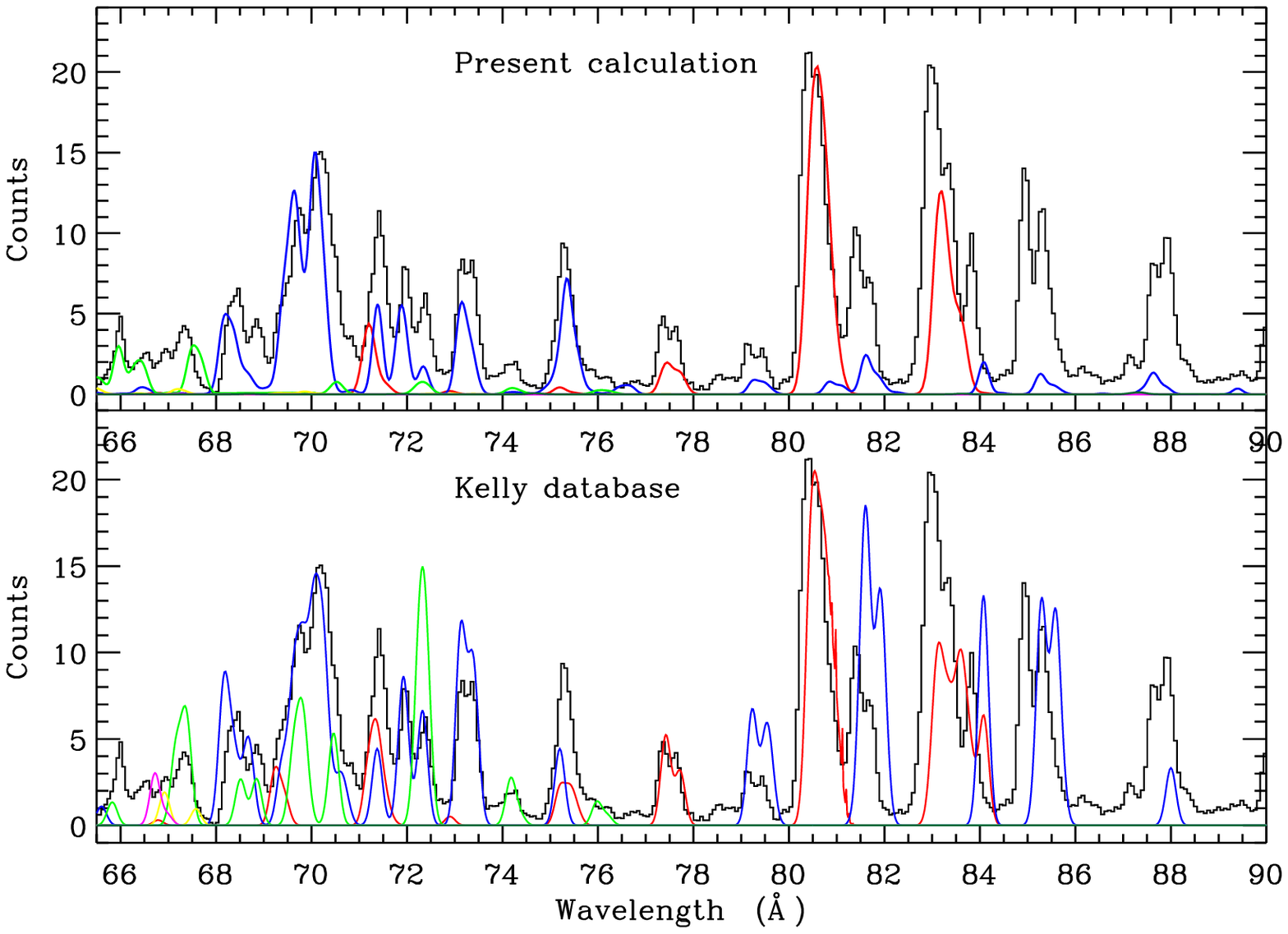}
\caption[short title]{(Online color)~~Experimental (histogram
curve) and synthesized spectra (color) of highly charged silicon
ions at a density of 1.0$\times$10$^{20}$cm$^{-3}$. The
calculations are normalized to the measured ones according to
lines at 80.353~\AA\,, 71.425~\AA\,, 62.052~\AA\,, 55.569~\AA\,,
50.278~\AA\,, 47.670~\AA\, and 44.135~\AA\, for Si~VI--Si~XII,
respectively.}
\end{figure*}
\clearpage

%figure-5
\clearpage
\begin{figure*}[h]
\centering
\includegraphics[angle=0,width=10cm,clip]{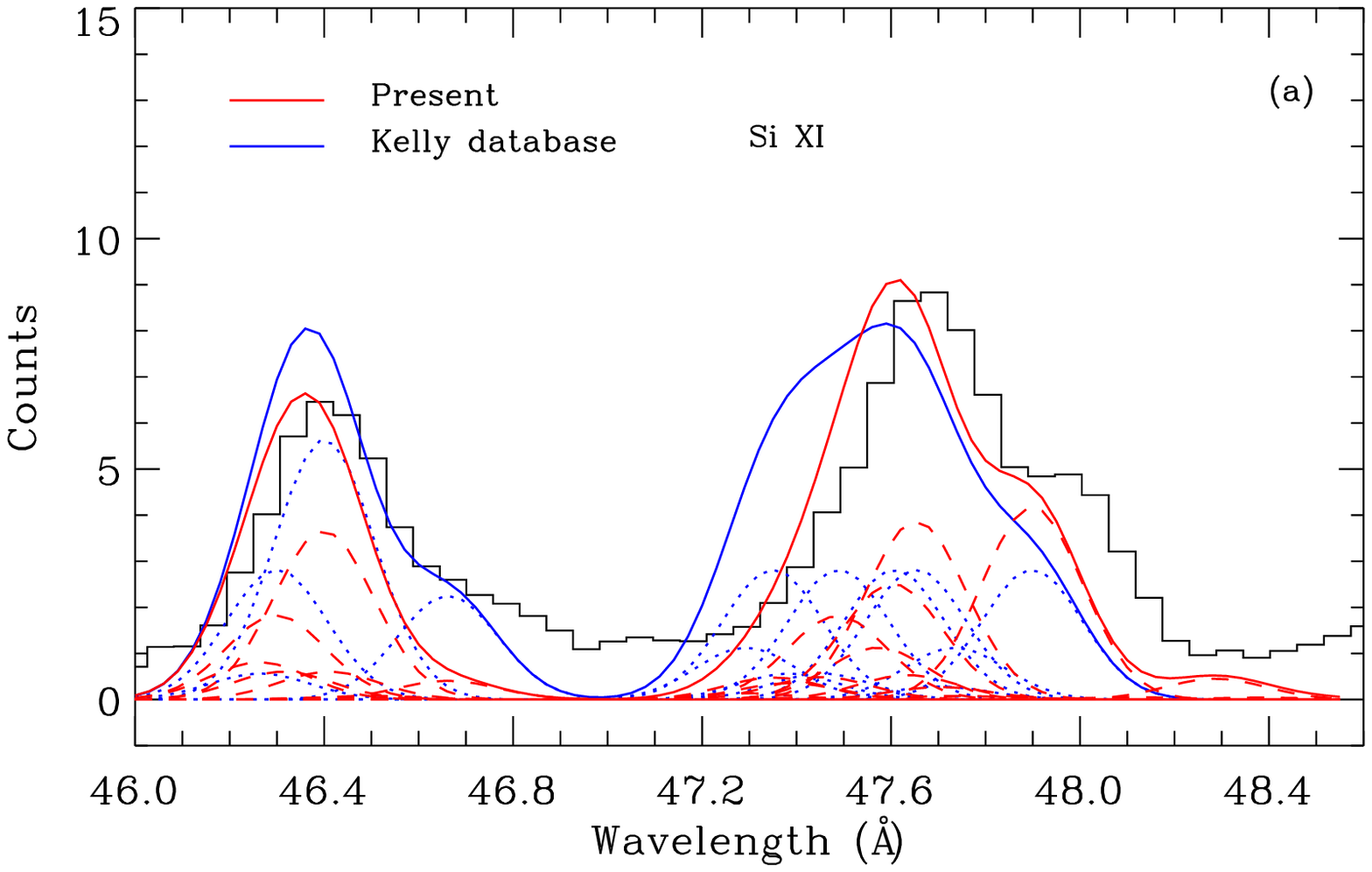} \\
\includegraphics[angle=0,width=10cm,clip]{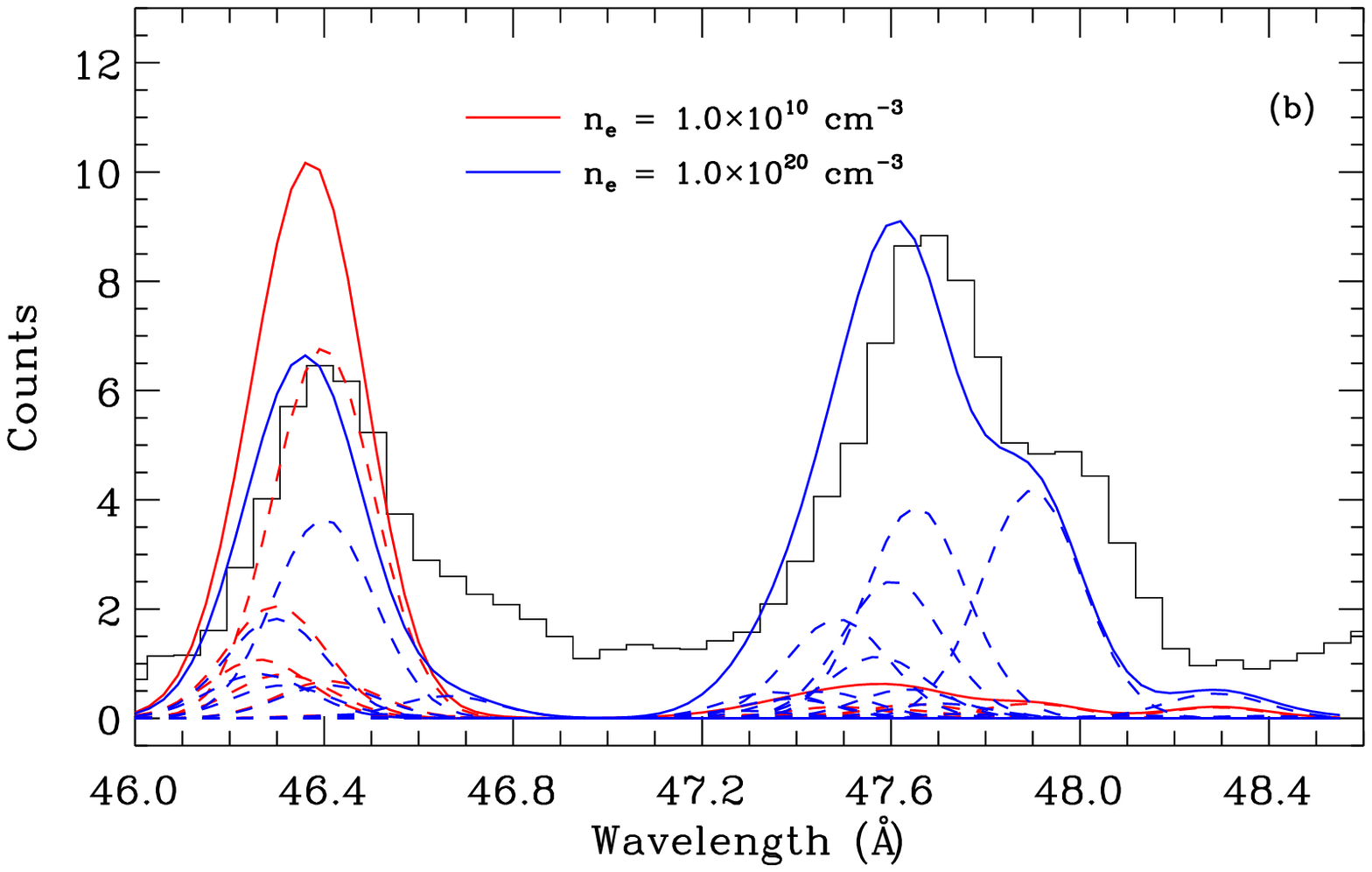}
\caption[short title]{(Online color)~~Emission lines of highly
charged Si~XI in range of 46--48.6~\AA\,. The theoretical line
intensities are folded by Gaussian profile with FWHM 0.3~\AA\,.
(a) Present calculated spectra of Si~XI (red-solid line added from
red-dashed lines for individual emissions of Si~XI), and the
spectra derived from Kelly database (blue-solid line added from
blue-dotted lines for individual emissions of Si~XI). The
theoretical spectra are normalized by the experimental peak at
47.67~\AA\,. (b) Present calculation for thin
(1.0$\times$10$^{10}$cm$^{-3}$, red-solid line added from
red-dashed lines for individual emissions of Si~XI) and dense
(1.0$\times$10$^{20}$cm$^{-3}$, blue-solid line added from
blue-dashed lines for individual emissions of Si~XI) plasmas.}
\end{figure*}
\clearpage

%figure-6
\clearpage
\begin{figure*}[h]
\centering
\includegraphics[angle=0,height=10cm,clip]{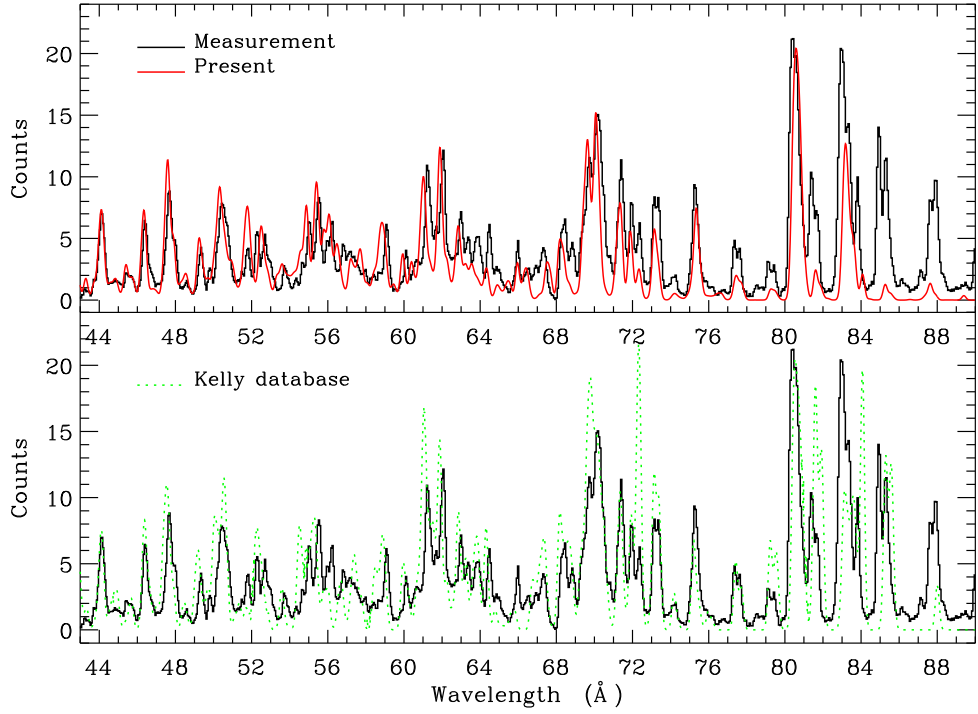}
\caption[short title]{(Online color)~~Experimental and theoretical
spectra of highly charged silicon ions in wavelength range of
43--90~\AA\,. Based upon the normalization [see caption in
Fig.~4], the sum of each charge state is shown by color curves for
present calculation (red-solid) and Kelly database
(green-dotted).}
\end{figure*}
\clearpage

\clearpage \pagestyle{empty}
\begin{deluxetable}{lccccccllllll}
\rotate \tablewidth{0pt} \tablecaption{Experimental wavelengths
along with statistical errors of 53 prominent emission lines in
wavelength range of 40---90~\AA\,, and their identifications. The
same number with different subscripts (in the Index column)
denotes the blending lines.} \tablehead{ \colhead{Index} &
\colhead{$\lambda_{\rm Exp}$}  & \colhead{Intensity} &
\colhead{$\lambda_{\rm Kelly}$} & \colhead{$\Delta\lambda$} &
\colhead{Ion} & \colhead{$\lambda_{\rm Pres.}$} &
\colhead{$\Delta\lambda$} & \colhead{Ion} &
\multicolumn{4}{c}{Transition}\\
\colhead{ }  & \colhead{\AA} & \colhead{Arb. Unit} & \colhead{\AA}
& \colhead{\AA} & \colhead{ } & \colhead{\AA} & \colhead{\AA} &
\colhead{ } & \multicolumn{2}{c}{Upper} &
\multicolumn{2}{c}{Lower} } \startdata
1   & 44.143$\pm$0.008  & 10.73$\pm$0.52 & 44.165 & -0.030 & Si~XII   & 44.165 & -0.022 & Si~XII  & $2s^23d         $ & $^2D_{5/2}$ & $2s^22p  $ & $^2P_{3/2}$\\
2a  & 45.440$\pm$0.038  & 3.31 $\pm$0.27 &        &        &          & 45.398 & 0.042  & Si~XI   & $2p3p           $ & $^1D_2    $ & $2s2p    $ & $^1P_1    $\\
2b  & 45.723$\pm$0.038  & 2.25 $\pm$0.28 & 45.684 & 0.039  & Si~X     & 45.756 & -0.033 & Si~X    & $2p^2(^1D)3p    $ & $^2P_{3/2}$ & $2s2p^2  $ & $^2D_{5/2}$\\
3   & 46.403$\pm$0.004  & 7.86 $\pm$0.21 & 46.401 & 0.002  & Si~XI    & 46.399 & 0.004  & Si~XI   & $2s3d           $ & $^3D_2    $ & $2s2p    $ & $^3P_2    $\\
4   & 47.670$\pm$0.006  & 10.52$\pm$0.33 & 47.653 & 0.017  & Si~XI    & 47.653 & 0.017  & Si~XI   & $2p3d           $ & $^3D_3    $ & $2p^2    $ & $^3P_2    $\\
5   & 47.989$\pm$0.008  & 4.78 $\pm$0.21 & 47.899 & 0.090  & Si~XI    & 47.899 & 0.090  & Si~XI   & $2p3d           $ & $^1F_3    $ & $2p^2    $ & $^1D_2    $\\
6   & 49.343$\pm$0.005  & 4.49 $\pm$0.16 & 49.222 & 0.121  & Si~XI    & 49.222 & 0.121  & Si~XI   & $2s3d           $ & $^1D_2    $ & $2s2p    $ & $^1P_1    $\\
7   & 49.828$\pm$0.011  & 2.29 $\pm$0.15 & 49.701 & 0.127  & Si~X     &        &        &         & $2s2p(^1P)3d    $ & $^2F_{7/2}$ & $2s2p^2  $ & $^2D_{5/2}$\\
8a  & 50.278$\pm$0.033  & 3.91 $\pm$0.36 & 50.333 & -0.055 & Si~X     & 50.333 & -0.055 & Si~X    & $2s2p(^3P)3d    $ & $^4D_{7/2}$ & $2s2p^2  $ & $^4P_{5/2}$\\
8b  & 50.517$\pm$0.033  & 8.63 $\pm$0.39 & 50.524 & -0.007 & Si~XI    &        &        &         & $2p3s           $ & $^3P_2    $ & $2p^2    $ & $^3P_2    $\\
9   & 51.796$\pm$0.008  & 4.48 $\pm$0.21 &        &        &          & 51.821 & -0.045 & Si~X    & $2p^2(^1D)3d    $ & $^2F_{7/2}$ & $2p^3    $ & $^2D_{5/2}$\\
10a & 52.291$\pm$0.004  & 6.27$\pm$0.19  & 52.155 & 0.136  & Si~X     &        &        &         & $2s2p(^1P)3d    $ & $^2D_{5/2}$ & $2s2p^2  $ & $^2P_{3/2}$\\
10b &                   &                & 52.296 & -0.005 & Si~XI    &        &        &         & $2s3s           $ & $^1S_0    $ & $2s2p    $ & $^1P_1    $\\
11  & 52.687$\pm$0.009  & 5.33 $\pm$0.28 & 52.485 & 0.202  & Si~X     & 52.484 & 0.203  & Si~X    & $2s2p(^3P)3d    $ & $^2F_{7/2}$ & $2s2p^2  $ & $^2D_{5/2}$\\
12a & 53.736$\pm$0.010  & 1.94 $\pm$0.17 & 53.573 & 0.163  & Si~X     & 53.572 & 0.164  & Si~X    & $2s2p(^3P)3d    $ & $^2D_{5/2}$ & $2s2p^2  $ & $^2D_{3/2}$\\
12b &       ~           &      ~         & 53.595 & 0.141  & Si~X     & 53.596 &        & Si~X    & $2s2p(^3P)3d    $ & $^2D_{3/2}$ & $2s2p^2  $ & $^2D_{3/2}$\\
13  & 54.705$\pm$0.011  & 2.30 $\pm$0.19 & 54.599 & 0.106  & Si~X     & 54.599 & 0.106  & Si~X    & $2s2p(^3P)3d    $ & $^2P_{3/2}$ & $2s2p^2  $ & $^2S_{1/2}$\\
14  & 55.034$\pm$0.004  & 6.19 $\pm$0.20 & 54.907 & 0.127  & Si IX    & 54.921 & 0.113  & Si~IX   & $2s2p^2(^4P)3d  $ & $^5D_3    $ & $2s2p^3  $ & $^5S_2    $\\
15  & 55.379$\pm$0.017  & 3.85 $\pm$0.63 & 55.272 & 0.107  & Si IX    & 55.272 & 0.107  & Si~IX   & $2s^22p3d       $ & $^3P_2    $ & $2s^22p^2$ & $^3P_2    $\\
16  & 55.569$\pm$0.010  & 6.71 $\pm$0.63 & 55.401 & 0.168  & Si IX    & 55.401 & 0.168  & Si~IX   & $2s^22p3d       $ & $^3D_3    $ & $2s^22p^2$ & $^3P_2    $\\
17  & 55.950$\pm$0.009  & 3.78 $\pm$0.21 & 55.781 & 0.169  & Si IX    & 55.779 & 0.171  & Si~IX   & $2s2p^2(^4P)3d  $ & $^5D_3    $ & $2s2p^3  $ & $^5S_2    $\\
18  & 56.229$\pm$0.006  & 5.69 $\pm$0.22 & 56.027 & 0.202  & Si~IX    & 56.027 & 0.202  & Si~IX   & $2s^22p3d       $ & $^1F_3    $ & $2s^22p^2$ & $^1D_2    $\\
19a & 59.088$\pm$0.003  & 6.74 $\pm$0.16 &        &        &          & 58.906 & 0.182  & Si~IX   & $2s2p^2(^4P)3d  $ & $^3F_4    $ & $2s2p^3  $ & $^3D_3    $\\
19b &                   &                & 59.004 & 0.084  & Si~IX    & 59.004 & 0.084  & Si~IX   & $2s2p^2(^4P)3d  $ & $^3F_3    $ & $2s2p^3  $ & $^3D_2    $\\
19c &                   &                & 58.885 & 0.203  & Si~VIII  &        &        &         & $2s2p^3(^5S)3p  $ & $^4P_{5/2}$ & $2s^22p^3$ & $^4S_{3/2}$\\
20  & 61.256$\pm$0.006  & 11.35$\pm$0.49 & 61.070 & 0.186  & Si VIII  & 61.032 & 0.224  & Si~VIII & $2s^22p^2(^1D)3d$ & $^2F_{5/2}$ & $2s^22p^3$ & $^4S_{3/2}$\\
21  & 62.052$\pm$0.004  & 14.88$\pm$0.33 & 61.914 & 0.138  & Si VIII  & 61.792 & 0.260  & Si~VIII & $2s^22p^2(^1D)3d$ & $^2F_{7/2}$ & $2s^22p^3$ & $^2D_{5/2}$\\
22  & 62.993$\pm$0.008  & 7.66 $\pm$0.33 & 62.849 & 0.144  & Si VIII  & 62.849 & 0.144  & Si~VIII & $2s^22p^2(^3P)4s$ & $^2P_{1/2}$ & $2s2p^4  $ & $^4P_{5/2}$\\
23  & 63.377$\pm$0.011  & 5.24 $\pm$0.29 & 63.229 & 0.148  & Si VIII  & 63.241 & 0.136  & Si~VIII & $2s2p^3(^3S)3p  $ & $^4P_{1/2}$ & $2s2p^4  $ & $^4P_{5/2}$\\
24  & 63.762$\pm$0.012  & 4.23 $\pm$0.33 & 63.732 & 0.030  & Si VIII  & 63.722 & 0.040  & Si~VIII & $2s^22p^2(^3P)3p$ & $^4P_{1/2}$ & $2s^22p^3$ & $^2D_{5/2}$\\
25  & 64.482$\pm$0.007  & 6.60 $\pm$0.27 & 64.327 & 0.155  & Si VIII  & 64.317 & 0.165  & Si~VIII & $2s^22p^2(^3P)3d$ & $^2D_{5/2}$ & $2s^22p^3$ & $^2P_{3/2}$\\
26  & 65.977$\pm$0.007  & 4.48 $\pm$0.25 & 65.833 & 0.144  & Si VIII  & 65.969 & 0.008  & Si~VIII & $2s2p^3(^3D)3d  $ & $^4G_{7/2}$ & $2s2p^4  $ & $^2D_{5/2}$\\
27  & 67.418$\pm$0.020  & 3.53 $\pm$0.41 & 67.408 & 0.010  & Si VIII  & 67.479 & -0.061 & Si~VIII & $2s2p^3(^3D)3d  $ & $^2G_{7/2}$ & $2s2p^4  $ & $^4P_{5/2}$\\
28a & 68.294$\pm$0.048  & 5.20 $\pm$3.01 & 68.148 & 0.146  & Si~VII   & 68.148 & 0.146  & Si~VII  & $2s^22p^3(^4S)3d$ & $^5D_3    $ & $2s^22p^4$ & $^3P_1    $\\
28b & 68.863$\pm$0.016  & 5.75 $\pm$0.47 & 68.715 & 0.148  & Si~VII   &        &        &         & $2s^22p^3(^2P)3d$ & $^3P_1    $ & $2s^22p^4$ & $^3P_0    $\\
29  & 69.734$\pm$0.007  & 15.37$\pm$0.48 & 69.663 & 0.071  & Si~VII   & 69.664 & 0.070  & Si~VII  & $2s^22p^3(^2D)3d$ & $^3F_2    $ & $2s^22p^4$ & $^3P_2    $\\
30a & 70.129$\pm$0.009  & 19.06$\pm$0.90 & 70.027 & 0.102  & Si~VII   & 70.072 & 0.057  & Si~VII  & $2s^22p^3(^2P)3d$ & $^3F_3    $ & $2s^22p^4$ & $^1D_2    $\\
30b &                   &                & 70.072 & 0.057  & Si~VII   &        &        &         & $2s^22p^3(^2P)3d$ & $^1F_3    $ & $2s^22p^4$ & $^1D_2    $\\
31a & 70.422$\pm$0.016  & 11.28$\pm$0.85 & 70.222 & 0.200  & Si~VII   & 70.185 & 0.237  & Si~VII  & $2s^22p^3(^2D)3d$ & $^3D_2    $ & $2s^22p^4$ & $^3P_1    $\\
31b &                   &                & 70.250 & 0.172  & Si~VII   &        &        &         & $2s^22p^3(^2P)3d$ & $^1P_1    $ & $2s^22p^4$ & $^1D_2    $\\
32  & 71.168$\pm$0.056  & 3.28 $\pm$1.06 & 71.181 & -0.013 & Si~VI    & 71.181 & -0.013 & Si~VI   & $2s^22p^4(^3P)4d$ & $^2D_{5/2}$ & $2s^22p^5$ & $^2P_{3/2}$\\
33  & 71.425$\pm$0.012  & 13.93$\pm$1.16 & 71.384 & 0.041  & Si~VII   & 71.384 & 0.041  & Si~VII  & $2s^22p^3(^2D)3d$ & $^3G_3    $ & $2s^22p^4$ & $^1D_2    $\\
34a & 71.953$\pm$0.008  & 9.57 $\pm$0.40 & 71.900 & 0.053  & Si~VI    & 71.842 & 0.111  & Si~VII  & $2s^22p^3(^2D)3d$ & $^3P_1    $ & $2s^22p^4$ & $^1D_2    $\\
34b &                   &                & 71.955 & -0.002 & Si~VI    & 71.955 & -0.002 & Si~VII  & $2s^22p^3(^2D)3d$ & $^1D_2    $ & $2s^22p^4$ & $^1D_2    $\\
35  & 72.389$\pm$0.010  & 7.24 $\pm$0.39 & 72.324 & 0.065  & Si~VII   & 72.324 & 0.065  & Si~VII  & $2s^22p^3(^2D)3d$ & $^3D_1    $ & $2s^22p^4$ & $^1D_2    $\\
36  & 73.136$\pm$0.004  & 7.50 $\pm$0.23 & 73.123 & 0.013  & Si~VII   & 73.123 & 0.013  & Si~VII  & $2s^22p^3(^2D)3d$ & $^3D_3    $ & $2s^22p^4$ & $^3P_2    $\\
37  & 73.394$\pm$0.004  & 7.44 $\pm$0.22 & 73.350 & 0.044  & Si~VII   & 73.311 & 0.083  & Si~VII  & $2s^22p^3(^2D)3d$ & $^3P_2    $ & $2s^22p^4$ & $^3P_1    $\\
38a & 75.160$\pm$0.038  & 8.48 $\pm$0.33 & 75.193 & -0.033 & Si~VII   & 75.335 & -0.175 & Si~VII  & $2s2p^4(^2P)3d  $ & $^1F_3    $ & $2s2p^5  $ & $^3P_2    $\\
38b & 75.352$\pm$0.042  & 3.87 $\pm$0.34 & 75.398 & -0.046 & Si~VI    &        &        &         & \\
39  & 77.348$\pm$0.006  & 4.43 $\pm$0.20 & 77.429 & -0.081 & Si~VI    & 77.429 & -0.081 & Si~VI   & $2s^22p^4(^1S)3d$ & $^2D_{5/2}$ & $2s^22p^5$ & $^2P_{3/2}$\\
40  & 77.626$\pm$0.007  & 3.74 $\pm$0.20 & 77.718 & -0.092 & Si~VI    & 77.718 & -0.092 & Si~VI   & $2s^22p^4(^1S)3d$ & $^2D_{3/2}$ & $2s^22p^5$ & $^2P_{1/2}$\\
41  & 79.132$\pm$0.030  & 3.26 $\pm$0.58 & 79.236 & -0.104 & Si~VII   & 79.237 & -0.105 & Si~VII  & $2s^22p^3(^4S)3s$ & $^5S_2    $ & $2s^22p^4$ & $^3P_2    $\\
42  & 79.448$\pm$0.035  & 2.73 $\pm$0.58 & 79.491 & -0.043 & Si~VII   & 79.444 & 0.004  & Si~VII  & $2s^22p^3(^4S)3s$ & $^5S_2    $ & $2s^22p^4$ & $^3P_1    $\\
43  & 80.353$\pm$0.006  & 23.54$\pm$1.03 & 80.449 & -0.096 & Si~VI    & 80.449 & -0.096 & Si~VI   & $2s^22p^4(^3P)3d$ & $^4D_{5/2}$ & $2s^22p^5$ & $^2P_{3/2}$\\
44  & 80.622$\pm$0.013  & 17.86$\pm$0.83 & 80.725 & -0.103 & Si~VI    & 80.725 & -0.103 & Si~VI   & $2s^22p^4(^3P)3d$ & $^4P_{3/2}$ & $2s^22p^5$ & $^2P_{1/2}$\\
45  & 81.382$\pm$0.010  & 11.75$\pm$0.60 & 81.620 & -0.238 & Si~VII   & 81.617 & -0.235 & Si~VII  & $2s^22p^3(^2D)3s$ & $^3D_3    $ & $2s^22p^4$ & $^3P_2    $\\
46  & 81.698$\pm$0.014  & 7.05 $\pm$0.61 & 81.895 & -0.197 & Si~VII   & 81.845 & -0.147 & Si~VII  & $2s^22p^3(^2D)3s$ & $^3D_2    $ & $2s^22p^4$ & $^3P_1    $\\
47  & 82.961$\pm$0.003  & 26.43$\pm$0.55 & 83.128 & -0.167 & Si~VI    & 83.128 & -0.167 & Si~VI   & $2s^22p^4(^3P)3d$ & $^2D_{5/2}$ & $2s^22p^5$ & $^2P_{3/2}$\\
48a & 83.328$\pm$0.005  & 17.19$\pm$0.52 & 83.611 & -0.283 & Si~VI    & 83.611 & -0.283 & Si~VI   & $2s^22p^4(^1D)3d$ & $^2D_{3/2}$ & $2s^22p^5$ & $^2P_{1/2}$\\
48b &                   &                & 83.526 & -0.298 & Si~VI    &        &   ~~   &         & $2s^22p^4(^3P)3d$ & $^4P_{5/2}$ & $2s^22p^5$ & $^2P_{3/2}$\\
49a & 83.809$\pm$0.007  & 10.46$\pm$0.48 & 84.082 & -0.273 & Si~VI    &        &   ~~   &   ~     & $2s^22p^4(^3P)3d$ & $^4F_{3/2}$ & $2s^22p^5$ & $^2P_{1/2}$\\
49b &                   &                & 84.082 & -0.273 & Si~VII   & 84.071 & -0.262 & Si~VII  & $2s^22p^3(^2D)3s$ & $^1D_2    $ & $2s^22p^4$ & $^1D_2    $\\
50  & 84.935$\pm$0.002  & 13.08$\pm$0.26 & 85.219 & -0.284 & Si~VII   &        &   ~~   &   ~     & $2s^22p^3(^2P)3s$ & $^1P_1    $ & $2s^22p^4$ & $^1S_0    $\\
51  & 85.330$\pm$0.004  & 10.68$\pm$0.30 & 85.584 & -0.254 & Si~VII   &        &   ~~   &   ~     & $2s^22p^3(^4S)3s$ & $^3S_1    $ & $2s^22p^4$ & $^3P_1    $\\
52  & 87.629$\pm$0.004  & 7.42 $\pm$0.26 & 88.008 & -0.379 & Si~VII   & 87.641 & -0.012 & Si~VII  & $2s2p^4(^4P)3s$   & $^3P_2    $ & $2s2p^5$   & $^3P_2    $\\
53  & 87.934$\pm$0.004  & 9.20 $\pm$0.27 & 88.330 & -0.396 & Si~XII(2)&        &   ~    &   ~     & $2s^23d         $ & $^2D_{5/2}$ & $2s^22p  $ & $^2P_{3/2}$\\
\enddata
\end{deluxetable}
\clearpage \pagestyle{plaintop}

\end{document}